\documentstyle[11pt]{article}

\def\Journal#1#2#3#4{{#1} {\bf #2}, #3 (#4)}

\def\NPB{{\em Nucl. Phys.} B}
\def\PLB{{\em Phys. Lett.}  B}

\def\PRD{{\em Phys. Rev.} D}

\def\PRT{\em Phys. Rept. }
\def\xxx#1           {{\sf hep-th/#1} }

\def\be{\begin{equation}}
\def\ee{\end{equation}}
\def\bea{\begin{eqnarray}}
\def\eea{\end{eqnarray}}
\newcommand{\D}{\Delta}

\begin{document}

\begin{center}
\null\vskip-24pt \hfill AEI-2001-107
\vskip 0.4truecm {\bf
ON EXCEPTIONAL NON-RENORMALIZATION PROPERTIES OF 
${\cal N}=4$ SYM$_4$}
\\
\vskip 0.5truecm
{\bf
G. Arutyunov
}
\\
\vskip 0.4truecm
{\it Max-Planck-Institut f\"ur Gravitationsphysik,
Albert-Einstein-Institut, \\
Am M\"uhlenberg 1, D-14476 Golm, Germany\\
Email:{\tt agleb@aei-potsdam.mpg.de}
}
\end{center}

\vskip 0.5truecm
\centerline{\bf Abstract} 
We discuss non-renormalization properties of some composite 
operators in ${\cal N}=4$ supersymmetric 
Yang-Mills theory.

\vskip 1cm

Recently considerable attention was attracted to the ${\cal N}=4$ super Yang-Mills 
theory basically due to the prominent role it plays among the models 
realizing the holographic AdS/CFT duality \cite{AGMOO}.
In the superconformal phase the dynamics of the gauge theory 
is encoded in the correlation functions of the composite gauge invariant operators,
which might exhibit in general a non-trivial behavior under the RG flow. 
In particular it is of great interest to determine the 4-point correlation 
functions (both holographic and weak coupling) 
of the ${\cal N}=4$ supercurrent (stress-tensor) multiplet $L$
and its OPE; the latter contains an information about many 
other composite operators present in the theory. 

A superconformal primary operator generating $L$ is a scalar $O^I$ of dimension 2
transforming in the irrep $\bf 20$ of the $R$-symmetry group $SU(4)$, $I=1,\ldots, 20$.
Presently both the holographic \cite{ArutFrol} and the weak coupling \cite{Pert} 
4-point correlators of $O^I$ and their OPE studies \cite{AFP,AEPS,AES} are available.\footnote{For
studies of other correlation functions from stress-tensor multiplet see e.g. \cite{DHMMR}.} 
Surprisingly composite operators\footnote{They saturate the bound of the so-called series A) 
of unitary irreps of $SU(2,2|4)$ and transform non-trivially under R-symmetry \cite{FZ}.}
with vanishing anomalous dimensions were found \cite{AFP} though naively unitarity allows 
the latter to appear in quantum interacting theory.   

This note is based on the paper \cite{AES} and reviews a statement  
that the OPE of two primary operators from the multiplet $L$ can contain superconformal
primary operators with a non-vanishing anomalous dimension {\it only} in the
singlet of $SU(4)$. 

It was found non-perturbatively \cite{EPSS} that 
the ``quantum'' part of the four-point function of $O^I$
comprising all possible quantum corrections to the free-field result is given
by a {\it single} function $F(v,u)$ of conformal cross-ratios, which we
choose to be  $v=\frac{x_{12}^2x_{34}^2}{x_{14}^2x_{23}^2}$ and
$u=1-\frac{x_{13}^2x_{24}^2}{x_{14}^2x_{23}^2}$. Under $SU(4)$ the product of two
$O^I$ decomposes as
$
{\bf 20}\times {\bf 20}={\bf 1}+{\bf 20}+{\bf 105}+{\bf 84}+{\bf 15}+{\bf 175} \, .
$
The ``quantum'' part of the four-point function of the operators $O^I$
projected on different irreps is 
\bea \label{proj}
\langle O(x_1)O(x_2)O(x_3)O(x_4)\rangle _i
=\frac{1}{x_{12}^4x_{34}^4}P_{i}(v,u) \frac{vF(v,u)}{(1-u)^2} \, , \eea 
where $P_{i}(v,u)$ are certain polynomials \cite{AFP,AEPS}.
Every irrep $i$ of $SU(4)$ in the OPE of two $O^I$ represents a contribution
from an infinite tower of operators $O_{\D,l}^{i}$, where $\D$ is the conformal
dimension of the operator, $l$ is its Lorentz spin. The corresponding
contribution to the four-point function can then be represented as an expansion
of the type
\begin{equation}
\label{cpwae}
\langle O(x_1)
O(x_2)O(x_3)O(x_4)\rangle_{i}
=\sum_{\Delta, l} a_{\Delta, l}^{i}{\cal H}_{\Delta, l}(x_{1,2,3,4})\; .
\end{equation}
Here ${\cal H}_{\Delta, l}(x_{1,2,3,4})$ denotes the (canonically normalized)
Conformal Partial Wave Amplitude (CPWA) for the exchange of an operator
$O_{\D,l}^{i}$ and $a_{\Delta, l}^{i}$ is a normalization constant. We treat
the CPWA as a double series of the type 
\begin{equation}
\label{gcpwa} {\cal H}_{\Delta, l} = \frac{1}{x_{12}^4 \, x_{34}^4} \,
v^{\frac{h}{2}} \, \sum_{n,m=0}^\infty c^{\Delta, l}_{nm} v^n u^m \, ,
\end{equation}
where the dimension $\D$ was split into a canonical part $\D_0$ and an
anomalous part $h$: $\D=\D_0+h$.
Assigning the grading parameter $T=2n+m$ to the monomial $v^nu^m$ one can show that
the monomials in (\ref{gcpwa}) with the lowest value of
$T$ have $T=\D_0$, where $\D_0$ is the canonical (free-field) dimension of the
corresponding operator.

Comparing (\ref{proj}) and (\ref{cpwae}) one finds, within every fractional
power $v^{\frac{h}{2}}$, the following compatibility conditions \bea
\label{comp} P_{i}\sum_{\Delta, l} a_{\Delta, l}^{j}{\cal
H}_{\Delta,l}(x_{1,2,3,4}) =P_{j}\sum_{\Delta, l} a_{\Delta, l}^{i}{\cal
H}_{\Delta,l}(x_{1,2,3,4}) \, \eea which hold for all pairs. 
Here the sums are taken over operators which have the same
$h$. Thus, eqs. (\ref{comp}) imply non-trivial relations between the CPWAs of
primary operators belonging to the same supersymmetry multiplet(s) with
anomalous dimension $h$.
Only one of these primary operators
is the superconformal primary operator, i.e., it generates under supersymmetry
the whole multiplet, while the others are its descendents.

Now we see that a superconformal primary operator appears only in the
singlet of $SU(4)$. Indeed, let us choose in (\ref{comp}) the irrep $j$ to be
the singlet. The polynomial $P_{\bf 1}$  is distinguished from the other
$P_i$'s by the presence of a constant term. Suppose that a superconformal
primary operator with a canonical dimension $\D_0$ contributes to the OPE and
transforms in some irrep $i$ which is not a singlet. Due to the constant in
$P_{\bf 1}$, the lowest-order monomials on the r.h.s. of (\ref{comp}) would
have $T=2n+m=\D_0$. Clearly, all the other $P_i$'s always raise the $T$-grading
by at least unity. The lowest dimension operator with canonical dimension
$\D_0'$ in the singlet would have the lowest terms with at least $T=\D_0-1$ (or
lower) to saturate (\ref{comp}). Hence, $\D_0'$ is always lower then $\D_0$,
and therefore the corresponding operator cannot be a supersymmetry descendent
of an operator in the irrep $i$. This shows that anomalous superconformal
primary operators are occure in the singlet of the $R$-symmetry group.


\end{document}